\documentclass{PoS}

\newcommand\ba{\begin{eqnarray}}
\newcommand\ea{\end{eqnarray}}
\newcommand\nn{\nonumber}
\newcommand{\be}{\begin{equation}}
\newcommand{\ee}{\end{equation}}
\newcommand{\M} {{\cal M}}

\title{Light meson emission in (anti)proton induced reactions}

\ShortTitle{Light meson emission}

\author{E.A. Kuraev\\
             JINR-BLTP, 141980 Dubna, Moscow region, Russian Federation }

\author{E.S. Kokoulina\\
             JINR-VBLHE, 141980 Dubna, Moscow region, Russian Federation }

\author{\speaker{E. Tomasi-Gustafsson}\\
 CEA,IRFU,SPhN, Saclay, 91191 Gif-sur-Yvette Cedex, France and\\
and Univ. Paris-Sud, IPNO, UMR-8608, Orsay, F-91405, France \\
        E-mail: \email{etomasi@cea.fr}}

\abstract{Reactions induced by high energy antiprotons on proton on nuclei are accompanied  with large probability by the emission of a few mesons.
Interesting phenomena can be observed and QCD tests can be performed, through the detection of one or more mesons. 

The collinear emission from high energy (anti)proton beams of a hard pion or vector meson,  can be calculated similarly to the emission of a hard photon from an electron \cite{Kuraev:2013izz}. This is a well known process in QED, and it is called the "Quasi-Real Electron method", where the incident particle is an electron and a hard photon is emitted leaving an 'almost on shell' electron impinging on the target \cite{Baier:1973ms}. 

Such process is well known as Initial State Emission (ISR) method of scanning over incident energy, and can be used, in the hadron case, to produce different kind of particles in similar kinematical conditions. 

In case of emission of a charged light meson, $\pi$ or $\rho$-meson, in proton-proton(anti-proton) collisions, the meson can be deviated in a magnetic field and detected.

The collinear emission (along the beam direction) of a charged meson may be used to produce high energy (anti)neutron beams. This can be very useful to measure the difference of the cross sections of (anti)proton and (anti)neutron scattering from the target and may open the way for checking sum rules with antiparticles.  Hard meson emission allows also to enhance the cross section when the energy loss from one of the incident particles lowers the total energy up to the mass of a resonance.

The cross section can be calculated, on the basis of factorized formulas, where the probability of emission of the light mesons multiplies the cross section of the sub-process. Multiplicity distributions for neutral and charged meson production are also given.
}

\FullConference{XXII International Baldin Seminar on High Energy Physics Problems,\\
		15-20 September 2014\\
		JINR, Dubna, Russia}

\begin{document}

\section{Introduction}

Among (anti)proton induced reactions which occur with high probability (the cross section is of the order of [mb]), we focus here to light meson production in the kinematical domain accessible at the future FAIR facility by the PANDA experiment (the antiproton momentum is in the range 1.5-15 GeV). 

\begin{figure}
\centering
\includegraphics[width=8cm]{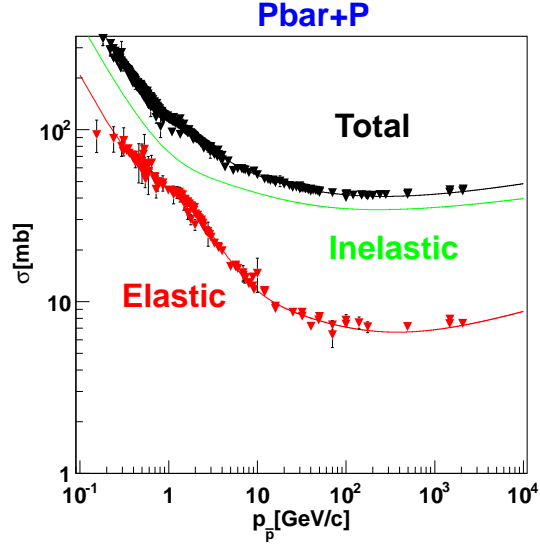}
\caption {Total (black) and elastic (red) cross section for the $\bar p +p$ reaction, as function of antiproton beam momentum $p_{\bar p}$. The contribution of inelastic events is also shown (green line).}
\label{fig:total}
\end{figure}

The world data for the total and elastic cross sections for $\bar p+p$ induced reactions are illustrated in Fig. \ref{fig:total} as function of  the laboratory antiproton  momentum \cite {Dbeyssi:2012zz}. From the figure we see that 
$\sigma_{\bar p p}^{el}/\sigma_{\bar p p}^{tot}\simeq 1/5$ and that the difference between the total and elastic cross
section parametrizations (green line), which represents the contribution of the inelastic events, is of the order of $40$~mb in the range accessible at PANDA.

Figs. 2 shows that the most probable reaction involving pions, corresponds to more than three pions in the final state. We parametrize the cross section for the reactions given in Fig. 2 as:
Among the inelastic processes, the charge exchange process $p\bar p \to n\bar n$ has the largest cross section:

\begin{figure}
\centering

\includegraphics[width=150 mm]{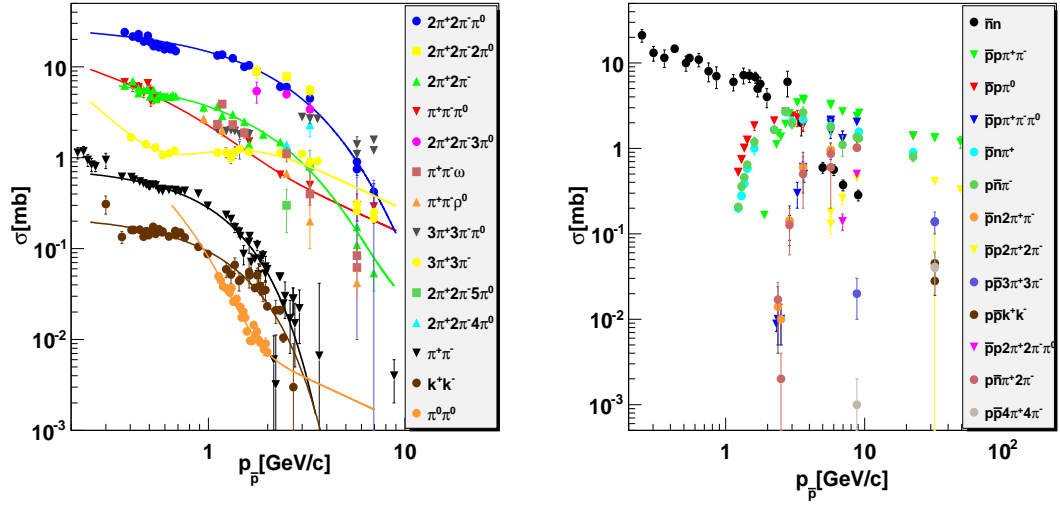}
\caption{Cross sections for different inelastic channels 
in the $\bar p +p$ reaction as a function of $p_{\bar p}$: (left) mesonic channels; (right) mesonic and hadronic channels }
\label{Fig:fig2}
\end{figure}

It is known from QED that two Feynman diagrams contribute to the process of Bhabha-scattering:
the scattering ($t$) and the annihilation ($s$) channels. Neglecting the interference  one can estimate the  relative contribution of the scattering diagram $|\M_s|^2$ and the annihilation one $|\M_a|^2$ for the case when the final particles are emitted at large angles in the system of center of mass:
\be
    \frac{ \sum |\M_a|^2 }{ \sum |\M_s|^2 } \approx \frac{1}{10},
\ee    
The contribution of scattering mechanism dominates. This statement holds even if hadron form factors are taken into account,
i.e., one can assume that a similar relative contributions of these two mechanisms would take place in case of elastic and inelastic $\bar pp \to \bar pp$ cross section. The processes relevant to "scattering" mechanism are denoted as peripheral processes. 

In this work we estimate the CE contribution, using the formalism \cite{BFK, Arbuzov:2010zza}. The emission by the initial proton of a charged light meson-$\pi$ or $\rho$-meson in proton-proton(anti-proton) collisions transforms high energy protons (for example in a proton beam) into neutrons. This effect is observed in accelerator physics \cite{Nikitin}. Note that hard collinear photon emission was also studied in Ref. \cite{Caffo:1984jb}. This works uses a frame where the relevant amplitude interferes with the background, i.e., all other emission mechanisms, except emission from the electron line, whereas \cite{BFK} applies the structure function approach (that we use). Although the methods are different, the answers of these publications are equivalent for the observables.

CE reactions may occur in antiproton beams, too. High energy, high intensity antiproton beams will be available in next future at PANDA \cite{PANDA}, FAIR \cite{FAIR}. Hard $\pi$ and $\rho$ meson can be detected with high efficiency. Charged meson will be deviated by the 2T magnetic field of the central spectrometer, and (anti)neutrons, which are produced at high rate, could be used as a secondary high energy beam. We are interested here in 'initial state emission', as a mechanism to produce a (anti)neutron beam  with comparable energy of the (anti)proton beam.

In this work, we propose a description of CE reactions referring to the known QED process of emission of a hard real photons by electron (positron) beams at $e^+e^-$ colliders. Such process enhances the cross section when the energy loss from one of the incident particles lowers the total energy up to  the mass of a resonance. This is known as "return to resonance" mechanism. In the case of creation of a narrow resonance this mechanism appears through a radiative tail: it is the characteristic behavior of the cross section which gradually decreases for energies exceeding the resonance mass. This mechanism provides, indeed, an effective method for studying narrow resonances like $J/\Psi$.

For the emission in a narrow cone along the directions of the initial (final) particles, the emission probability has a logarithmic enhancement, which increases with the energy of the "parent" charged particle. In frame of QED this mechanism is called as "quasi-real electron" mechanism (QRE) \cite{BFK}.

In this work we apply the QRE mechanism to the case of hadrons and, in particular, to the collinear emission of a light meson from a (anti)proton beam. We evaluate the cross section for this process for single as well as multi pion production, where pions can be neutral or charged. Our derivation concerns a  kinematical region outside the resonance formation region.
%%%%%%%%%%%%%%%%%%%%%%%%%%%%%%%%%%%%%%%%%%%%%%%%%%%%%%%%%%%%%%%%%%%%%%%%%
\subsection{Quasi-real electron kinematics with hard photon emission}
%%%%%%%%%%%%%%%%%%%%%%%%%%%%%%%%%%%%%%%%%%%%%%%%%%%%%%%%%%%%%%%%%%%%%%%%%

\begin{figure}
\begin{center}
\includegraphics[width=8cm]{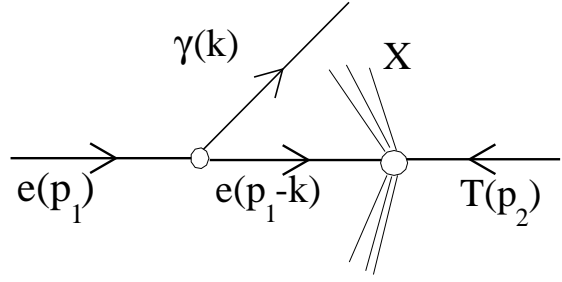}
\caption{Feynman diagram for collinear hard photon emission in $e T$ reactions (T stands for any target)}
\label{Fig:Fig1}
\end{center}
\end{figure}

Let us consider the radiative process  $e(p_1)+T(p_2) \to e(p_1-k) +\gamma (k) + X$ (the four momenta of the particles are written in parenthesis), $T$ stay for any nuclear target(proton $p$, or nucleus $A$), and the final state $X$ is undetected, Fig. \ref{Fig:Fig1}.

The virtual electron after the hard (collinear) photon emission is almost on mass shell \cite{BFK}. This property allows to express the matrix element of the radiative process $e(p)+T \to e(p-k) +\gamma(k)+X$ in terms of the matrix element of the non-radiative process $e+T \to e+X$:
\be
\M_\gamma(p_1,p_2)=e\bar{\cal T}(p_2)\frac{\hat{p_1}-\hat{k}+m}{-2p_1k}\hat{\varepsilon}(k)u(p_1).
\label{eq:eq1}
\ee

In the case when the denominator of the intermediate electron's Green function is small
 $|(p_1-k)^2-m^2|\ll2p_1p_2$ one can write $\hat{p}_1-\hat{k}+m=\sum_s u^s(p_1-k)\bar{u}^s(p_1-k)$ and the matrix element has a factorized form.

The square of the matrix element, summed over the spin states of the photon is:
\be
\sum |\M_\gamma|^2=e^2\left [\frac{E_p^2+E_{\vec{p}-\vec{k}}^2}{\omega(E_p-\omega)(k p)}-\frac{m^2}{(k p)^2}\right ]\sum|\bar{\cal T}(p_2)u(p_1-k)|^2.
\label{eq:eq2}
\ee
where $\sum|\bar{\cal T}(p_2)u(p_1-k)|^2$ is the Born matrix element squared with shifted argument.

In the case of unpolarized particles, the cross section of the process $e(p_1)+T(p_2)\to e+\gamma+T$ may be written in factorized form:
\ba
d\sigma_\gamma(s,x)&=&d\sigma(\bar{x}s)dW_\gamma(x), \bar{x}=1-x, \nn \\
dW_\gamma(x)&=&\frac{\alpha}{\pi}\frac{d x}{x}\left [(1-x+\frac{1}{2}x^2)\ln\frac{E^2\theta_0^2}{m_e^2}-(1-x)\right ],~ x=\frac{\omega}{E}, ~\theta<\theta_0\ll 1, ~\frac{E\theta_0}{m_e}\gg 1,
\label{eq:eq3}
\ea
where $E$ is the energy of the initial electron (center of mass frame implied $\vec{p}_1+\vec{p}_2=0$), and $ s=(p_1+p_2)^2$.

It is assumed here that the initial electron transforms into an electron with energy fraction $1-x$ and a hard photon with
energy fraction $x$ which is emitted within the cone $\theta<\theta_0$ along the direction of initial electron.
Moreover it is implied that $\bar{x}s>s_{thr}$, where $s_{thr}$ is the threshold energy of process without photon emission.
The logarithmic enhancement originates from the small values of the mass of the intermediate electron, which is almost on mass shell. This justifies the name of Quasi Real Electron (QRE) method.

Below we consider a possible extension of the QRE method to the processes with "quasi real" (anti)nucleon intermediate state.

%%%%%%%%%%%%%%%%%%%%%%%%%%%%%%%%%%%%%%%%
\subsection{Application to hadron physics}
%%%%%%%%%%%%%%%%%%%%%%%%%%%%%%%%%%%%%%%%

Let us apply this formalism to the case of initial high energy proton (anti-proton) beams and the emission of a hard pion or vector meson in forward direction, collinear to the beam. We do not consider the emission of the pion from the final proton (final state emission), which corresponds to a different kinematical region for the intermediate particle.

\begin{figure}
\begin{center}
\includegraphics[width=9cm]{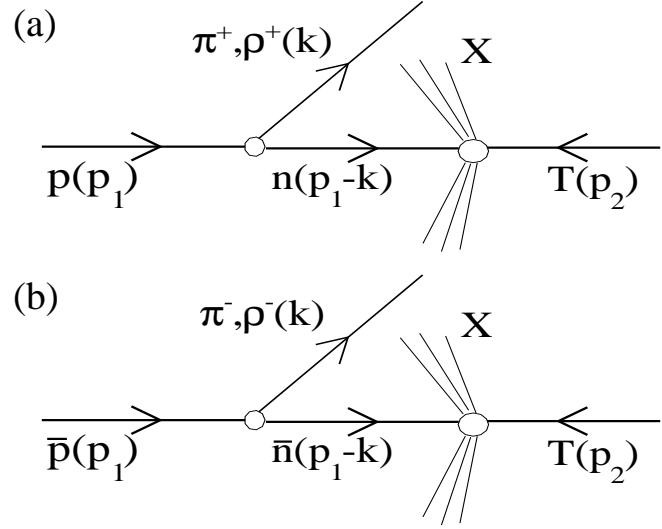}
\caption{Feynman diagram for collinear hard charged pion ($\rho$-meson) emission in $p(\bar p)+T$ collisions.}
\label{Fig:Fig2}
\end{center}
\end{figure}

For the case of emission of a positive-charged $\rho^+$ or $\pi^+$ meson by the high energy (anti)proton, the final state consists in a high energy (anti)neutron, accompanied by a positively charged meson. The charged meson can be deflected by an external magnetic field, providing the possibility to select a high energy neutron beam. In the case of emission of the neutral meson, it can be identified by measuring its decay channels.

Let us consider the reactions (Fig. \ref{Fig:Fig2}):
\ba
& p + T&\to n + T +h^+ \\
&\bar p + T& \to \bar n + T +h^-
\ea
where $h=\rho$ or $\pi$ and $T$ may be any target ($p$, $n$, nucleus..). In QED the interaction is assumed to occur through the exchange of a virtual photon. In the present case the particle can be a vector meson. However, the nature of the exchanged particle is irrelevant for the present considerations, where we focus on small angle charged pion emission and on the factorization properties of the cross section.

The matrix element for collinear $\pi (\rho)$ emission can be written as:
\ba
\M^{pT}_{h_+}(p_1,p_2))&=& \M_{nT}(p_1-k,p_2){\cal T}^{pn}_{h_+}(p_1,p_1-k), \nn\\
\M^{\bar{p}T}_{h_-}(p_1,p_2))&=& \M_{\bar{n}T}(p_1-k,p_2){\cal T}^{\bar{p}\bar{n}}_{h_-}(p_1,p_1-k), 
\label{eq:Mch}
\ea
where
\ba
{\cal T}_{\pi}^{pn}(p_1,p_1-k)&=&\frac{g}{m_h^2-2p_1k}\bar u_n(p_1-k)\gamma_5u_p(p_1),\nn\\
{\cal T}_{\rho}^{pn}(p_1,p_1-k)&=&\frac{g}{m_h^2-2p_1k}\bar u_n(p_1-k)\hat\epsilon u_p(p_1).
\ea
are the matrix elements of the subprocesses: $p\to n+\pi^+$ and $p\to n+\rho^+$ (or
$\bar p\to \bar n+\pi^-$ and $p\to n+\rho^-$). The matrix element of emission of a charged pion is expressed in factorized form (\ref{eq:Mch}) of a term connected to the target and a 'universal' factor for the hard meson emission, which generalizes the QED result. Below, we will focus on this last term.

The relevant cross sections are:
\ba
d\sigma^{pT\to h_+X}(s,x)&=&\sigma^{nT\to X}(\bar{x}s)d W_{h_+}(x), \nn \\
d\sigma^{\bar{p}T\to h_+X}(s,x)&=&\sigma^{\bar{n}T\to X}(\bar{x}s)d W_{h_-}(x), \nn \\
d\sigma^{pT\to h_0X}(s,x)&=&\sigma^{pT\to X}(\bar{x}s)d W_{h_0}(x).
\ea

The quantity $d W_{\rho}(x)$ can be inferred using the QED result:
\ba
\frac{dW_{\rho^i}(x)}{dx}&=&\frac{g^2}{4\pi^2}
\frac{1}{x}\sqrt{1-\frac{m_\rho^2}{x^2E^2}}\left [\left (1-x+\frac{1}{2}x^2\right )L-(1-x)\right],
\nn \\
1>x&=&\frac{E_\rho}{E}>\frac{m_\rho}{E}, ~
L= \ln\left (1+\frac{E^2\theta_0^2}{M^2}\right ), \rho^i=\rho^+,\rho^-,\rho^0,
\label{eq:eqrho}
\ea
where $M$, $ m_\rho$, $E$, and $E_\rho$ are the masses and the energies of the initial proton and the emitted $\rho$-meson (Laboratory reference frame implied).

\begin{figure}
\begin{center}
\includegraphics[width=12cm]{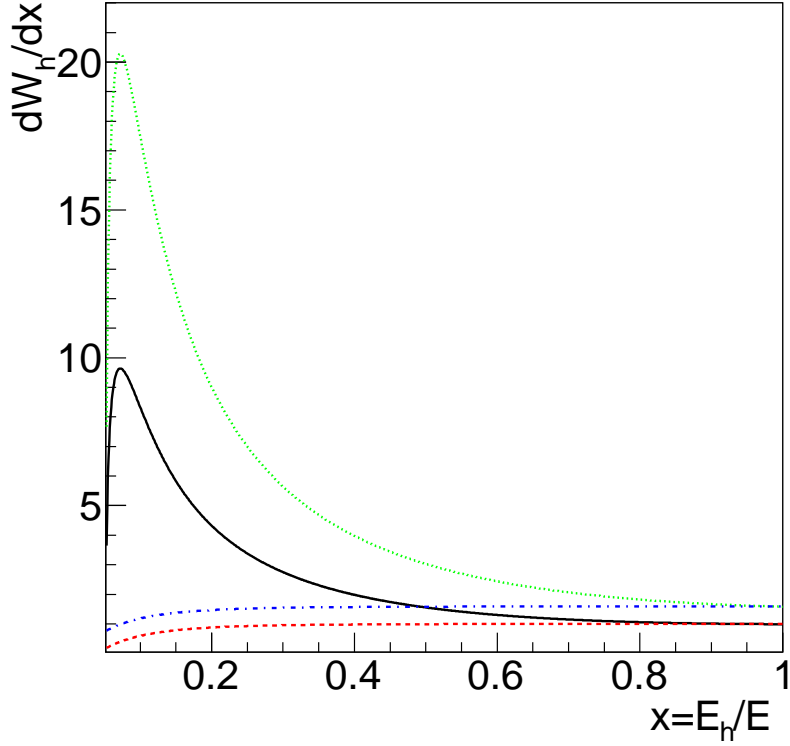}
\caption{Color online. Quantity $dW_h/dx$ for $\rho$- and $\pi$-meson production as a function of the meson energy fraction $x=E_{h}/E$, from Eqs. (\protect\ref{eq:eqrho},\protect\ref{eq:eqpi}), for $E = 15$ GeV and for two values of the $h$-emission angle: $\theta_0=10^{\circ}$ for $\rho$- meson (black, solid) and for $\pi$-meson (red, dashed),
$\theta_0=20^{\circ}$  for $\rho$- meson (green, dotted) and for $\pi$-meson (blue, dash-dotted).}
\label{Fig:hadronprob}
\end{center}
\end{figure}

For the probability of hard pion emission we have
\ba
\frac{d W_\pi}{d x}=\sum|{\cal T}_{pn}(p_1,p_1-k)|^2\frac{d^3 k}{16\omega\pi^3},
\ea
with
\ba
\sum|{\cal T}^{pn}(p_1,p_1-k)|^2=\frac{g^2}{[m_\pi^2-2(p_1k)]^2}Tr (\hat{p}_1-\hat{k}+M)\gamma_5(\hat{p}_1+M)\gamma_5= \nn \\
\frac{4(p_1k)g^2}{\left [m_\pi^2-2(p_1k)\right ]^2}, (p_1k)=E\omega(1-b c), 1-b^2\approx\frac{m_\pi^2}{\omega^2}+\frac{M^2}{E^2},
\label{eq:eqTpn}
\ea
with $c=\cos(\vec{k},\vec{p}_1)$. The angular integration in the region $1-(\theta_0^2/2)<c<1$ leads to
\ba
&&\frac{dW_{\pi^i}(x)}{dx}=\frac{g^2}{8\pi^2}\sqrt{1-\frac{m_\pi^2}{x^2E^2}}\left [L+\ln\frac{1}{d(x)}+\frac{m_\pi^2}{x d(x)M^2}\right],
\nn \\
&&1>x=\frac{E_\pi}{E}>\frac{m_\pi}{E},~d(x)=1+\frac{m_\pi^2\bar{x}}{M^2x^2},~\bar{x}=1-x,
~\pi^i=\pi^+,\pi^-,\pi^0,
\label{eq:eqpi}
\ea
where  $g=g_{\rho p p}=g_{\pi p p}\approx 6$ is the strong coupling constant.The quantities $d W_h(x)/d x$ as functions of the energy fraction $x=E_h/E$ ($h=\rho,\pi$) are drawn in Fig. \ref{Fig:hadronprob}, for $E=15$ GeV and two different values of $\theta_0$: $10^{\circ}$ and $20^{\circ}$.

The expressions of the integrated probabilities are:
\ba
W_h&=&\int\limits_{x^h_t}^1 \frac{d W_h}{d x}d x=\frac{g^2}{4\pi^2}(A^hL+B^h), \nn\\
A^\rho&=&I_0(x_t^\rho)-I_1(x_t^\rho)+\frac{1}{2}I_2(x_t^\rho),~
B^\rho=-I_0(x_t^\rho)+I_1(x_t^\rho),\nn\\
A^\pi&=&\frac{1}{2}I_1(x_t^\pi);~B^\pi=I_1(x_t^\pi),
\label{eq:eqpr}
\ea
where $x^h_t=E^h_{t}/E$ with $E^h_{t}=m^h$ - threshold value of the energy of the detected particle, $h=\rho,\pi$. The analytic expressions of the functions $I_i(z)$, $i=0,1,2$ are presented in Appendix.

The integrated quantities $W_h$, $h={\rho,\pi}$ can, in general, exceed unity, violating unitarity. To restore unitarity, we have to take into account virtual corrections: the vertex for the emission of a single pion (charged or neutral) from the proton has to include 'radiative corrections', which account for emission and absorption of any number of virtual pions. For this aim we use the known expression for the probability of emission of $n$ "soft" photons in processes of charged particles hard interaction, $i.e.,$ the Poisson  formula for emission of $n$ soft photons $W^n=(a^n/n!)e^{-a}$ (where $a$ is the probability of emission of a single soft photon) \cite{AkhBer81}. 

Note that the probability of emission of 'soft' neutral pions follows a Poisson distribution, which is not the case for the emission of charged pions. Fortunately, in our case, it is sufficient to consider the emission of one charged pion at lowest order (the process of one charged pion emission) plus any number of real and virtual pion with total charge zero. In such configuration, this vertex has the form of the product of the Born probability of emission of a single pion multiplied by the Poisson-like factor:
\be
P_{\pi,\rho}=e^{-W_{\pi,\rho}},
\label{eq:probh}
\ee
which takes into account virtual corrections.

\begin{figure}
\begin{center}
\includegraphics[width=7cm]{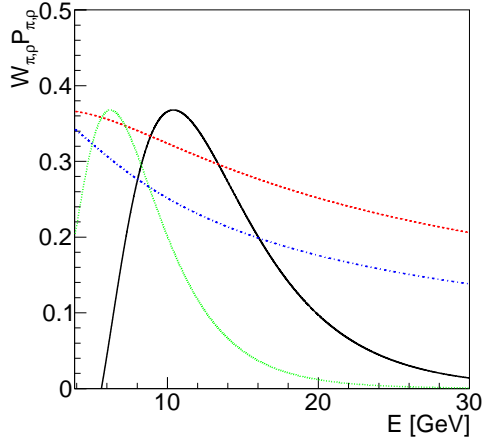}
\caption{Renormalized probabilities $W_{\rho,\pi}\cdot P_{\rho,\pi}$, as function of the incident energy for two values of the hadron emission angle. }
%Notations as in Fig. \ref{Fig:hadronprob}.}
\label{Fig:intprob}
\end{center}
\end{figure}

The renormalized probabilities $W_{\pi,\rho}P_{\pi,\rho} <1$ from Eqs. (\ref{eq:eqpr},\ref{eq:probh}) are illustrated in
Fig. \ref{Fig:intprob} as a function of energy, for two different $\theta_0$ angles.

\begin{figure}
\begin{center}
\includegraphics[width=8cm]{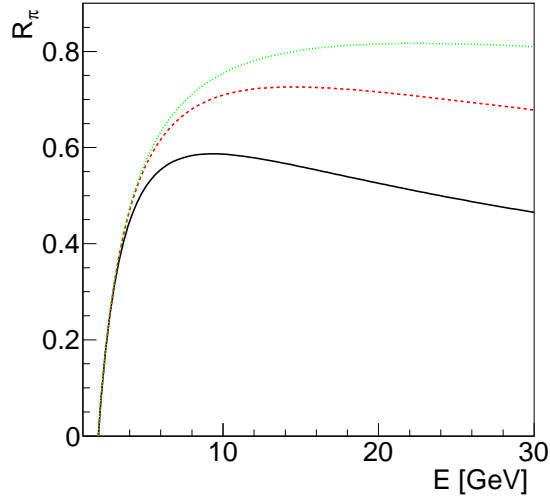}
\caption{Renormalization factor $R_{\pi}$, 
%Eq. (\ref{eq:eqppi}), 
for the probability of emission of 2 (black, solid line), 3 (red, dashed line), 4 (green, dotted line) neutral pions, as a function of the incident energy for $\theta_0=10^{\circ}$.}
\label{Fig:ppi}
\end{center}
\end{figure}

Keeping in mind the possible processes of emission of $n$ real soft neutral pion escaping the detection, the final result can be obtained using the replacement
\be
\sigma(s) \to \sigma(s)\times {\cal R_{\pi}},~{\cal R_{\pi}}=P_\pi\sum_{k=0}^{k=n}\frac{W^k_\pi}{k!}.
\label{eq:eqppi}
\ee
The renormalization factor ${\cal R}_\pi$ is illustrated in Fig. \ref{Fig:ppi}, for the probability of emission of 2 (black, solid line), 3 (red, dashed line), 4 (green, dotted line) pions.

The quantity $P_\pi$ can be compared to the experimentally measurable phenomena \cite{Nikitin}: the fraction of protons in the final state of
proton-proton collisions is approximately one half, $P_\pi\approx 0.5$. The commonly accepted explanation is that charge exchange reactions are responsible for changing protons into neutrons.

Let us consider the antiproton-proton annihilation into two and three  pions $\bar p+p\to \pi^+\pi^-$ or $\bar p+p\to \pi^+\pi^-\pi^0$.

Concerning the production of two charged pions, accompanied by a final state $X$, we can write:
\be
d\sigma^{p\bar p\to \rho^0 X}= 2\frac {d W_\rho(x)}{dx}\sigma^{p\bar p\to X}(\bar{x}s)\times P_{\rho},~
\label{eq:rhocs}
\ee
where the factor of two takes into account two kinematical situations, corresponding to the emission along each of the initial particles and $P_{\rho}$ is the survival factor (\ref{eq:probh}) which takes into account virtual radiative corrections. The characteristic peak at $x=x_{max}$ has the same nature as for the QED process $e^+ + e^-\to \mu^+ +\mu^- +\gamma$. As explained in Ref. \cite{Baier}, it is a threshold effect, corresponding
to the creation of a muon pair, where $x_{max}=1-4M_\mu^2/s$, $M_{\mu}$ is the muon mass.

\begin{figure}
\begin{center}
\includegraphics[width=12cm]{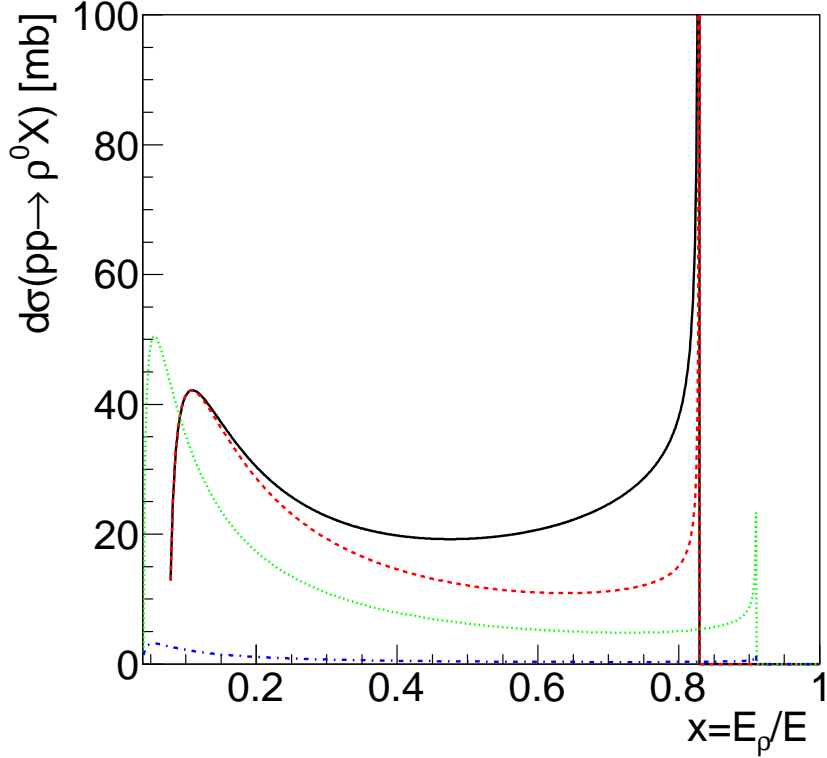}
\caption{The cross section $d\sigma(p,\bar p\to \rho^0 X)$ is plotted as function of the $\rho$ energy fraction for two values of the incident energy and of the $\rho$ emission angle: $E = 10$ GeV and $\theta_0=10^{\circ}$ (black, solid line),   $E = 10$ GeV and $\theta_0=20^{\circ}$ (red, dashed line), $E = 20$ GeV and $\theta_0=10^{\circ}$ (green, dotted line),   $E = 20$ GeV and $\theta_0=20^{\circ}$ (blue, dash-dotted line).}
\label{Fig:cs}
\end{center}
\end{figure}

The cross section (\ref{eq:rhocs}) is illustrated in Fig. \ref{Fig:cs} for two different values of the laboratory energy and of the emitted angle as a function of the $\rho$ meson energy fraction.

In case of three pion production,assuming that the process occurs through a $\pi^0\rho^0$ initial state emission, we find:
\ba
d\sigma(p,\bar{p})^{p\bar{p}\to \pi\rho X}&=&dW^0_\rho(x_\rho) dW^0_\pi(x_\pi)[d\sigma(p-p_\rho,\bar{p}-p_\pi)^{p\bar{p}\to X}+ \nn \\
&&d\sigma(p-p_\pi,\bar{p}-p_\rho)^{p\bar{p}\to X}]P_\pi P_\rho,
\ea
implying the subsequent decay $\rho^0 \to \pi^+\pi^-$.

It is interesting to note that the cross sections for the interaction of high energy neutron (anti-neutron) beams with a hadronic target can be calculated using the cross sections of proton beam interacting with the same target with the emission of the charged meson. We obtain (see Eqs. (\ref{eq:eqrho},\ref{eq:eqpi})):
\ba
\sigma^{nT\to X}(\bar{x}s)=\frac{d\sigma^{pT\to h^+ X}/d x}{d W_+(x)/d x},
\ea
and similarly for the anti-proton beams.

Experimental data from coincidence experiment with the selection of a hard meson at small angles are not available, and we can not test the factorisation prediction. However the values of the cross section for $\bar p p \rho^0 X$ illustrated in Fig. \ref{Fig:cs} are not in contradiction with the corresponding total cross sections (for a compilation, see \cite{Dbeyssi:2012zz}).

Inversely, if we use the experimental data for the total cross section of process $\bar{p}p \to \bar{n} h^-p\approx$ 1 mb,  we can predict the value of the total cross section  of process $\bar{n}p \to X$
\ba
P_\pi W_\pi(E_1,\theta_0)\sigma^{\bar{n}p\to X}(E-E_1) =\sigma^{\bar{p}p\to \pi X}(E),
\ea
with $W_\pi(E,\theta_0)$  given in Eqs. (\ref{eq:eqrho},\ref{eq:eqppi}).
%%%%%%%%%%%%%%%%%%%%%%%%%%%
\section{Conclusions}
%%%%%%%%%%%%%%%%%%%%%%%%%%%

We have extended the QRE method to light meson emission from an (anti)proton beam and calculated the probabilities and the relative cross section for multi-pion emission. The considered processes can be measured at present and planned hadron facilities. We have also suggested a possible application.  The collinear light meson emission could also be used to produce secondary (anti)neutron beams, at a high energy (anti)proton accelerator. This would constitute an alternative to the usual way, when high-energy neutron beams are produced as secondary beams, by break-up of  deuterons on a hadronic target.

Our result for the matrix element squared, Eq. (\ref{eq:eqTpn}), agrees with Adler principle: the cross section vanishes when the pion momenta vanishes. This result can be inferred as well from the reduction formula of current algebra \cite{Sakurai}. Note that the probabilities to create a $\pi$ or $\rho$-meson by a proton, can also be obtained using the infinite momentum reference frame, (Ref. \cite{Altarelli:1977zs}, Eq. (52)).

This allows one to obtain the relation between the cross section
$d\sigma_k(s(1-x))$ and $d\sigma_1(1)$. 
This relation has the form of conversion on the energy
fraction of the emitted particle $x=k_1^0/p_1^0$ of the probability of
emission by the initial projectile $dW_k(x)$ with cross section
$d\sigma_k(s(1-x))$. We underline that the details of interaction of projectile
with the target are irrelevant, assuming that they are the same for
shifted and unshifted kinematics.
The contribution of the final state hard hadron emission as well as the
interference of the relevant amplitude with the one describing the initial
state emission vanish in the limit $\theta_0 \to 0$.

Let us discuss the limits of our approach and the accuracy of the present calculation. Concerning the kinematical region where this mechanism is important, we consider charged pion emission at angles, smaller than an angle $\theta_0$, i.e. $\theta < \theta_0 \ll 1$.  Moreover $E \theta >> M$, which insures the presence of large logarithm. In this case, the main contribution is due to the large quantity $L = \ln( E^2 \theta^2 / M^2 ) \gg 1$. In the calculations, we keep the terms containing $L$ and omit the terms proportional to $\theta^2$. The resulting accuracy is given by $1 + O(1/L)$. Taking $L=10$, the accuracy of the calculation is of the order of 10\%.

We considered the case when the initial and the virtual
projectiles are fermions. For instance we used the density of probability
to find a fermion into the initial fermion (see 
(\ref{eq:eq2},\ref{eq:eq3}))
$p_e^e(z)=[1+(1-z)^2]/z$.
Instead one can consider other relations of the same kind, which hold 
when the initial projectile is a boson (photon, $\rho$, meson, gluon). 
We will not disuss this topic here.
Another interesting possibility is the conversion of the initial lepton to a photon
or a neutral $Z$ boson.
In this case we must use $P_e^\gamma(z) =z^2+(1-z)^2$ \cite{Baier}.

The arguments given above have a phenomenological character and are formulated in terms of hadrons. A similar idea, at quark level, was introduced in Ref. \cite{Teryaev82}, where the emission of $\rho$-meson by quark and the $\rho$ meson production in quark-antiquark annihilation was studied. Special attention was paid to polarization phenomena of the created $\rho$-meson.

We have also suggested a possible application.  The collinear light meson emission could also be used to produce secondary (anti)neutron beams, at a high energy (anti)proton accelerator. This would constitute an alternative to the usual way, when high-energy neutron beams are produced as secondary beams, by break-up of deuterons on a hadronic target.

In frame of the "Gluon Dominance Model", developed by one of us \cite{GDM} the ratio of the inelastic CE  cross section to the total inelastic cross section in $pp$ scattering is estimated as 40\%, in reasonable agreement with the experimental data \cite{Murzin}.

The collinear light meson emission mechanism in (anti)proton-proton collisions provide a possible source of events with rather high multiplicities of (charged and neutral) pion production. For such events the emission of hadrons in initial as well as in final states must be taken into account.

We considered here  only initial proton emission. In this case, the resonance formation is kinematically forbidden since the four momentum of the virtual nucleon is Space-Like. The simplest CE processes $p\bar{p} \to n \bar{n}$,~ $\bar{p} n \to \bar{\Lambda} \Sigma^-$ can be in principle measured at PANDA. Other reaction mechanisms can contribute to these processes. In the frame of a description in terms of a single pseudoscalar meson $\pi^+$, $K^+$ exchange, information on the strange meson-baryon constant can be extracted. Note that neutrons created by $\rho$ and $\pi$ emission form a (anti)proton beam with the mechanism considered here (initial state radiation) can not be formed by subsequent decay of a nucleon resonance, as $\Delta$. Nucleon resonances can not be produced in the initial state, because the neutron has the time like momentum squared (the propagator is $(p-k)^2-M^2<0$) . 

The simplest CE processes $p\bar{p} \to n \bar{n}$,~ $\bar{p} n \to \bar{\Lambda} \Sigma^-$ can be in principle measured at PANDA. Other reaction mechanisms can contribute to these processes. In the frame of a description in terms of a single pseudoscalar meson $\pi^+, K^+$ exchange, information on the strange meson-baryon constant can be extracted.
  
\section{Acknowledgements}

One of us $EAK$ is grateful to RFBR grant 11-02-00112 for support. We are grateful to S. Barkanova, A. Alexeev, and V. Zykunov, for interest to this problem and to V.A.~Nikitin for useful discussions.

\section{Appendix}

The analytic expressions for calculating the integrals
$$I_n(z)=\int\limits_z^1\frac{d x}{x}x^n\sqrt{1-\left(\frac{z}{x}\right )^2}$$
are:
\ba
I_0(z)&=&\frac{1}{2}\ln\frac{1+r}{1-r}-r; \nn \\
I_1(z)&=&r+z \arcsin (z); \nn \\
I_2(z)&=&\frac{1}{2}r-\frac{z^2}{4}\ln\frac{1+r}{1-r}; \nn
\ea

with  $r=\sqrt{1-z^2}$.

%%%%%%%%%%%%%%%%%%%%%%%%%%%
\section{Instead of Conclusions:  A memory of Professor Eduard Alekseevich KURAEV}
%%%%%%%%%%%%%%%%%%%%%%%%%%%

%\begin{center}
%\begin{figure}[h]
%\includegraphics[width=0.7\textwidth]{Photo_EAK.eps}
%\end{figure}

%Prof. Eduard Alekseevich KURAEV

%Krasnodar, 17/10/1940 ~~~~~~~~~~Dubna, 4/3/2014
%\end{center}
{\it
Prof. Eduard Alekseevich KURAEV passed away, while working on this subject. 

It is an invaluable loss for our community, in particular for the colleagues who had the chance to meet him and to share his encyclopedical knowledge, his enthusiasm, and his original and surprising ideas.

Always concentrated on solving difficult problems, He published hundreds of scientific papers, and collected thousands of citations.

He had an astonishing energy, extraordinary ability and resistance for effort and hard work., a passion for sharing ideas, discoveries, and results in seminars and in  teaching and educating young students and researchers. He was always ready to listen and discuss with the colleagues knocking at his door, coming from near or far Institutes.

He had a deep understanding and a curiosity in the observation of Nature, of the visible and invisible features of Nature, and, at the same time, a strong capability in abstraction and a passion for mathematical problems.

We loose with Prof. Eduard Alekseevich Kuraev, an engaging and unique personality, and an exceptional physicist.  
}

\newpage

\end{document}